\documentclass{article}

\usepackage{arxiv}

\usepackage[utf8]{inputenc} 
\usepackage[T1]{fontenc}    
\usepackage{hyperref}       
\usepackage{url}            
\usepackage{booktabs}       
\usepackage{amsfonts}       
\usepackage{nicefrac}       
\usepackage{microtype}      
\usepackage{lipsum}		
\usepackage{graphicx}
\usepackage{natbib}
\usepackage{doi}
\usepackage{xcolor}
\newcommand{\ER}[1]{\textcolor{black}{#1}}

\usepackage{gensymb}

\usepackage{adjustbox}
\usepackage{ulem}
\usepackage{cancel}
\usepackage{svg}
\usepackage{amsmath}
\newcommand{\de}{\mathrm{d}}
\usepackage{caption}
\usepackage{graphicx}
\usepackage{comment}
\DeclareCaptionLabelFormat{adja-page}{\hrulefill\\#1 #2 \emph{(previous page)}}

\title{Thickness profiles of giant soap films}

\author{ \href{https://orcid.org/0000-0003-0327-3605}{\includegraphics[scale=0.06]{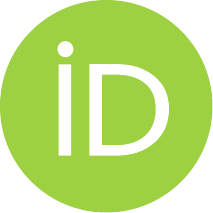}\hspace{1mm}Marina Pasquet$^1$,} \href{https://orcid.org/0000-0001-7803-6677}{\includegraphics[scale=0.06]{orcid.pdf}\hspace{1mm}Frédéric Restagno$^1$,} \href{https://orcid.org/0000-0002-3256-1254}{\includegraphics[scale=0.06]{orcid.pdf}\hspace{1mm}Isabelle Cantat$^2$,} \href{https://orcid.org/0000-0002-3201-6811}{\includegraphics[scale=0.06]{orcid.pdf}\hspace{1mm}Emmanuelle Rio$^1$,}\\
	$^1$ Universit\'e Paris-Saclay, CNRS, Laboratoire de Physique des Solides, 91405, Orsay, France.\\
	$^2$ Universit\'e de Rennes, CNRS, IPR (Institut de Physique de Rennes) - UMR 6251, 35000 Rennes, France.\\
}

\renewcommand{\shorttitle}{Generation of Giant Soap Films}

\hypersetup{
pdftitle={Generation of Giant Soap Films},
pdfauthor={Marina Pasquet, Frédéric Restagno, Isabelle Cantat, Emmanuelle Rio},
}

\begin{document}
\maketitle

\begin{abstract}
\ER{Production, drainage and stability of foams films, \textit{i.e.} films in contact with their menisci, are fascinating problems that remain still unsolved.  In this article, we propose to explore the regime of large velocities and large film sizes. This one is not accessible in experiments classically conducted in the literature, and allows us to study the regime of large extension and large extension rates. }
\ER{With our setup,} we make soap films up to two meters high by pulling a horizontal fishing line driven by belts out of a soapy solution at velocities ranging from 20~cm/s to 250~cm/s. 
We characterize the thickness profile of the central part of the film that behaves like a rubber band under tension.
We show that its thickness profile is well described by a static model in which a homogeneous elastic film is stretched by its own weight. 
This leads to an exponential thickness profile with a characteristic length given by a competition between gravity and surface elasticity. 
The prefactor is fixed by the shape and area of the film, governed by the fishing line motion but also by a continuous extraction of foam film from the lateral menisci, thicker than the central part, and that progressively  invades the film from its lateral boundaries. The model we propose captures the subtle interplay between gravity, film elasticity and film extraction and leads to predictions in good agreement with our experimental data.
\end{abstract}


\section{Introduction}
Bubble artists are able to create giant soap bubbles and films, \textit{i.e.} larger than one meter, on a daily basis in their artistic performances. One example is Graeme Denton's world record for the largest vertical soap bubble exceeding 10~m in height,  generated by pulling a ring out of a bath containing a soap solution at a velocity of about three meters per second \cite{WorldRecord}. 
In the artistic field, these records are regularly broken and this one dates from 2020. From a fundamental point of view, the existence of these giant objects raises many questions that are still to be elucidated, such as what limits this maximum size, the role of the entrainment velocity to make them, or even what sets the thickness profile of these giant films. Inspired by these ephemeral giant objects, we studied the generation of planar soap films up to two meters in size.
 
In the literature, soap films of comparable size are always fed from above \cite{Ballet2006, Kellay2002, Kellay2017, Rutgers2001, Sane2018, Salkin2016, Kim2017}.
Therefore we have developed a new experimental setup, presented in detail in a previous article \cite{mariot2021new}. The principle of the film generation is identical to what has been developed to study small soap films \cite{Lionti-Addad1992, Cohen-Addad1994, Adelizzi2004, Berg2005, Saulnier2011, Saulnier2014, Seiwert2017, Champougny2018} up to 10~cm: a frame made by a fishing line is pulled out of a bath at a controlled velocity and drags a soap film.
At the small velocities (up to 10 cm/s) usually used in such experiments, the thickness at the bottom of the films $h_{\text{Fr}}$ is given by a balance between viscous stress and capillarity and follows Frankel's law \cite{Mysels1959}: 
\begin{equation} \label{eq:loi_Frankel}
h_{\text{Fr}} \ =  \ 1,89 \ \ell_c\ \text{Ca}^{2/3}
\end{equation}
\noindent
where $\ell_c = \sqrt{\gamma_0/(\rho g)}$ is the capillary length and $\text{Ca} = \eta V/\gamma_0$ the capillary number. 
$\gamma_0$ is the surface tension of the solution, $\rho$ its density, $\eta$ its viscosity, $g$ the gravitational acceleration constant and $V$ the entrainment velocity. 
\ER{Such extracted  films, of thickness $h_{\text{Fr}}$,  will be called Frankel's films in the following.}
The good agreement between this law and experimental data has been studied extensively as well as the deviations observed with complex fluids \citep{Cohen-Addad1994}, at high velocities \citep{Berg2005,seiwert2014theoretical,champougny2015surfactant} or at the top of the film \citep{Saulnier2011}.

During the film generation, \ER{the film is stretched.} 
In the limit of poorly soluble surfactants, the creation of the film surface area leads to a diminution of the surfactant concentration $\Gamma$. 
Surface tension and surfactant surface concentration are linked by the Gibbs surface elasticity \cite{gibbs1948collected,Prins1967}, 
\begin{equation} \label{Eq_def_EGibbs}
E=- \frac{\partial \gamma}{\partial \ln \Gamma},
\end{equation}
which quantifies the surface tension variation subsequent to a change in surface concentration.
This results in a surface tension in the film higher than in the meniscus, where the tension is assumed to remain close to its equilibrium value $\gamma_0$.
The interfacial stress in the film, denoted $\gamma_0 + \Delta \gamma$, leads to the extraction of a film from the menisci, especially from the lateral ones \cite{Cantat2013, Seiwert2014}. The hydrodynamic laws in the vicinity of the meniscus impose a coupling between $ \Delta \gamma$, the extracted film thickness and the extraction velocity.  In the limit of high Gibbs modulus and in a steady state regime, this coupling is derived from the Frankel's law (Eq. \ref{eq:loi_Frankel}) \cite{Mysels1959,seiwert2014theoretical,Cantat2013}: 
\begin{equation} \label{eq:Frankel_deltagamma}
\Delta \gamma \ = 3.85 \ \gamma_0 \  \text{Ca}^{2/3} = 2.03 \ \gamma_0 \ \frac{h_{\text{Fr}}}{\ell_c}
\end{equation}
\ER{\noindent with $h_{\text{Fr}}$ given by Eq. \ref{eq:loi_Frankel}.}
A film element at a given altitude is stretched not only by the entrainment of the fishing line but also by the weight of the film underneath \citep{Lucassen1981,Couder1981,Gennes2001}. 
This leads to an additional vertical gradient of surface tension in the film and therefore to a gradient of extraction velocity along the vertical meniscus. 
The main goal of this article is to measure the thickness profile  in the central part of the film  during its generation, and to predict its evolution  with a model that takes into account these different  effects. 

In this article, we first  present in the part \ref{Part:setup} the setup used to generate  giant vertical films (up to 2 m) by pulling a fishing line out of a liquid bath at high velocity (between 20 and 250 cm/s).   
Our experimental results obtained by visualization and spectrometry will be described in part \ref{Part:experimental_characterization} and we will propose an \textit{in-situ} measurement of the surface tension gradient in a film, just after the motors stop.
The hydrostatic model, which predicts the film thickness profile and the surface tension gradient is proposed in the part \ref{Part:theoretical_description} together with a numerical resolution of the proposed model. A direct comparison with measurements will be presented at the end of the part \ref{part:comparaisonExpManip}.

\section{Presentation of the Experimental Setup} \label{Part:setup}


\begin{figure}[!ht]
  \centering
    \includegraphics[width=0.85\linewidth]{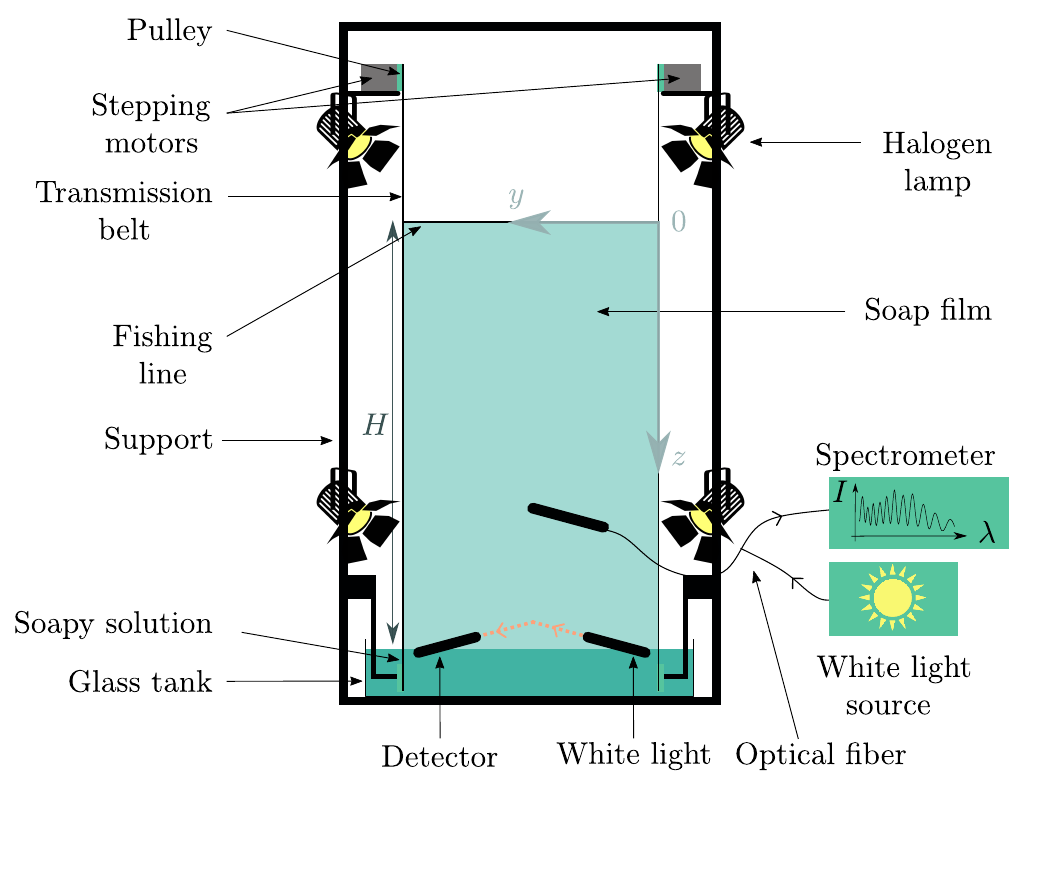}
 \caption{Schematic representation of the experimental setup allowing the generation of giant films. }
   \label{Fig:schema_setup_filmsgeants}
\end{figure}

The experimental setup for the generation of vertical soap films has been described previously \cite{mariot2021new} and we will only outline the main features here. 
The films are formed by pulling a fishing line (in fluorocarbon with a diameter of 0.74 mm) out of a glass container, as we can see in the Fig. \ref{Fig:schema_setup_filmsgeants}. 
This wire is driven out of the soap solution by two transmissions toothed belts. 
These belts are themselves driven by two coupled stepping motors (SM56 3 18 J4.6 from Rosier Mecatronique) associated with pulleys.
The motors were chosen in order to be able to control the velocity of the generation of the films between 20 and 250~cm/s. \ER{The acceleration takes place in the bath, before the fishing line reaches the surface and therefore the films are generated at constant velocity \cite{mariot2021new}.}
The maximum size of the films we can generate with this device is 2~m high and 70~cm wide. 

The setup allows automatic detection of the presence of the films, by measuring the reflection of a white light by three photoresistances. 
In addition, thickness measurements are also possible using a UV-VIS spectrometer (Ocean Optics Nanocalc 2000) associated with an optical fiber of 400 $\mu$m diameter. 
This spectrometer, placed at the center of the films in the $y$ direction (as defined in Fig. \ref{Fig:schema_setup_filmsgeants}), allows a measurement of the local thickness and can be moved on the whole height of the films. 
To control the humidity in the environment where the films are generated, we used two commercial humidifiers (one located at the bottom of the device, the other at the top of the chamber). 
The bottom humidifier (Okoia AH450) allows us to select the target humidity and the top humidifier (Bionaire BU1300W-I), whose flow rate is adjustable, allows us to have a more homogeneous humidity in the chamber, especially when we want to work in a very humid environment. 
In the following, the humidity is fixed at $85 \pm 4 $ \%, so we will neglect evaporation. 
The temperature is measured throughout the experiments with a typical uncertainty of 0.1 $^{\circ}$C and corresponds to the ambient temperature, which is about 22 $^{\circ}$C. 

To make direct visualizations of the soap films, we illuminate them thanks to a large white photo studio background paper (BD 129 Super White 2.72 × 11 m), which serves as a light reflector for four halogen lamps. We made a hole in it to fit the wide-angle lens (Nikon AF-P DX 10-20 mm) of a camera (Nikon D7200). 

For each experiment, we use 5 liters of soap solution made of 4 \% dishwashing liquid (Fairy from Procter \& Gamble, containing between 15 \% and 30 \% of anionic surfactants) with ultrapure water, with resistivity greater than 18.2 M$\Omega \cdot$cm. 

\section{Experimental characterization of giant soap films} \label{Part:experimental_characterization}

In this section, we present experimental observations and thickness measurement data obtained using the experimental setup presented in part \ref{Part:setup}. 

\subsection{Experimental observations} \label{part:direct_vizualisations}
\begin{figure}[!ht]
  \centering
    \includegraphics[width=0.9\linewidth]{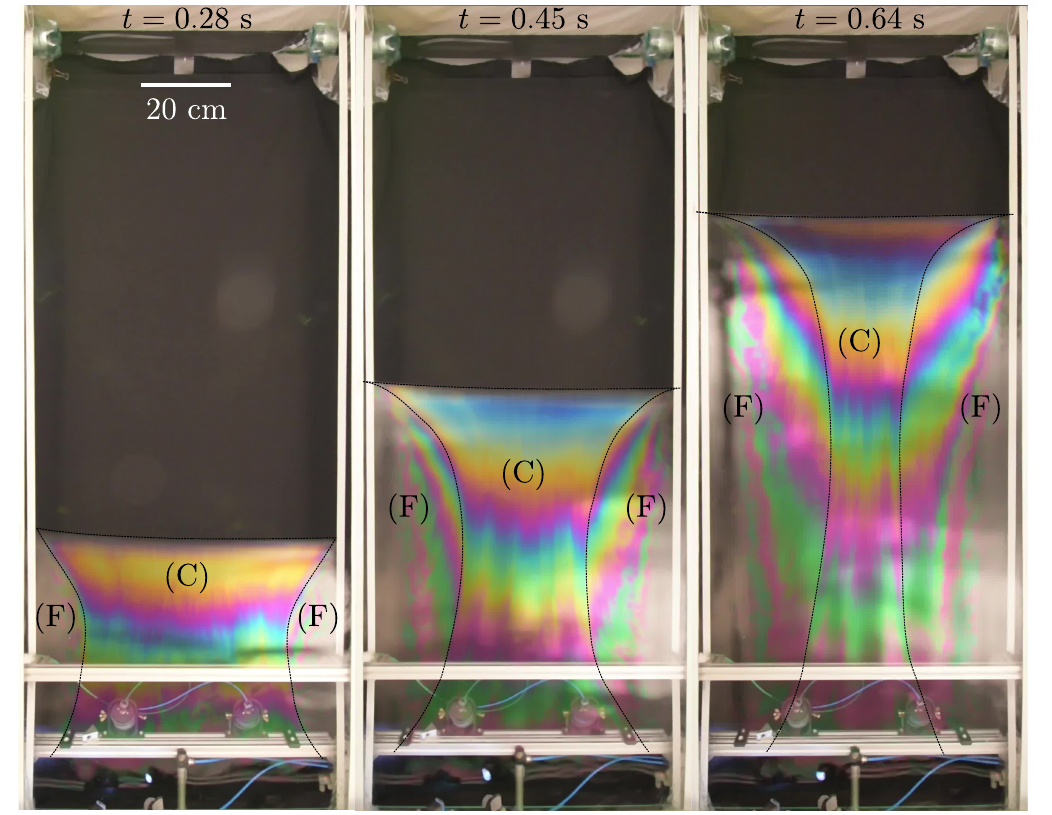}
 \caption{Photographs showing the evolution over time of a soap film generated at a fixed velocity of 200 cm/s. Thicker films extracted from the lateral menisci are observed: they are called Frankel films and are noted (F). The central film is noted (C). 
 }
   \label{Fig:photos_films-geants_montrant-zones-laterales}
\end{figure}

The qualitative behavior of the giant film during its generation is illustrated in Fig. \ref{Fig:photos_films-geants_montrant-zones-laterales}, the interferential colors providing us a direct indication of the film thickness.
We can see here that there is a central part (C) in which the film color is uniform at a given height, indicating a thickness stratification by the gravity. 
On the image on the left, on both sides of this central film we observe two lateral parts (F), which are thicker and not stratified.
They correspond to the Frankel's films, extracted from the lateral menisci. 
The drawing of the boundaries in Fig. \ref{Fig:photos_films-geants_montrant-zones-laterales} was done by hand, under ImageJ.
In the following, we will focus on the measure and prediction of the central film thickness profile. 
We will show that this necessitates a description of the lateral films extraction dynamics.

\subsection{Film thickness} \label{Part:subs_driven_thickness}

\begin{figure}[!ht]
  \centering
    \includegraphics[width=\linewidth]{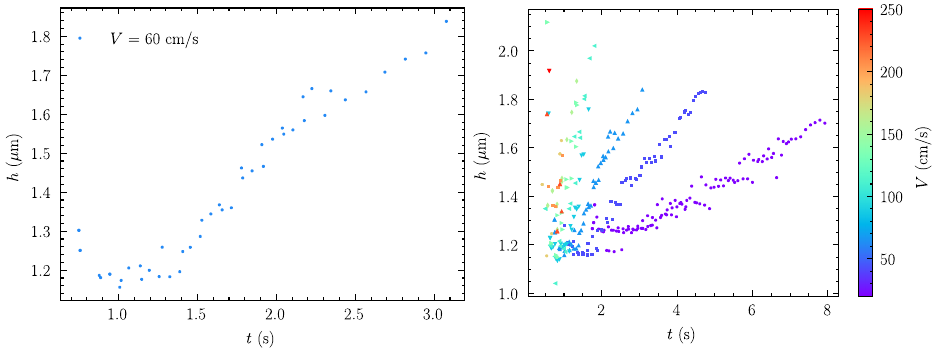}
   \caption{Evolution of the thickness  $h$ of soap films during the generation as a function of time $t$. These measurements were performed by spectrometry at the center of the films, at 13~cm above the surface of the bath. (a) Time evolution of the thickness for $V = 60$~cm/s. (b) Time evolution of the thickness for all the velocities probed between 20~cm/s (purple) and 250~cm/s (red). }
   \label{Fig_plot_spectro-RH85_position-h=f(t)-13cm_exv60cms-1}
\end{figure}

Fig. \ref{Fig_plot_spectro-RH85_position-h=f(t)-13cm_exv60cms-1} (a) shows the evolution of the thickness of a film entrained at a velocity of 60~cm/s as a function of time, for a spectrometer position fixed at 13~cm above the surface of the solution, at the center of the film (in the $y$ direction). 
The time $t = 0$ corresponds to the moment when the fishing line leaves the bath.
We  observe that up to $t \simeq 1$~s there is a thinning of the central part of the  film. 
After reaching a minimum value around 1.2~$\mu$m, the thickness $h$ increases linearly with time. 
This non-monotonic behavior is a robust phenomenon, as we can see in Fig.  \ref{Fig_plot_spectro-RH85_position-h=f(t)-13cm_exv60cms-1} (b), for all the measured velocities. We will show that this can be recovered by the model developed.

\begin{figure}[htbp]
  \centering
    \includegraphics[width=0.9\linewidth]{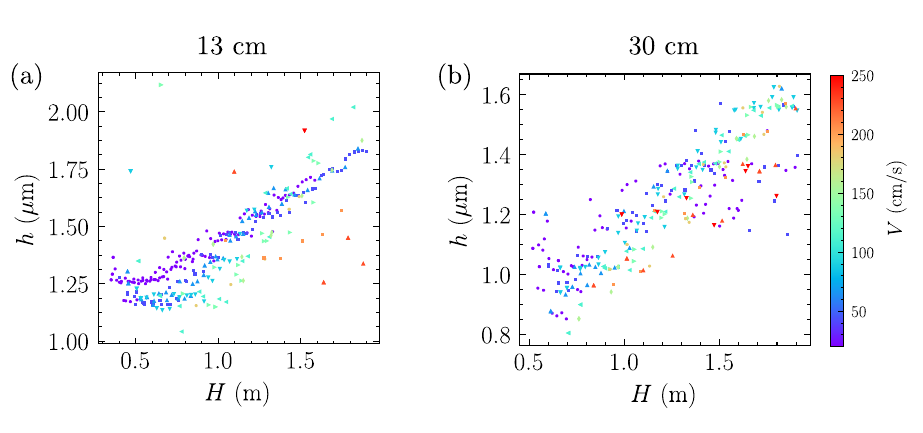}
 \caption{Evolution of the thickness $h$ of the soap films during the generation as a function of the height $H$ reached by the films, for all the velocities probed between 20~cm/s and 250~cm/s. These measurements are made by spectrometry (in the center of the films in the $y$ direction) at 13~cm above the solution surface (a) and at 30~cm above the surface (b). In the figure (a), this is the same data as shown in Fig. \ref{Fig_plot_spectro-RH85_position-h=f(t)-13cm_exv60cms-1} (b). With this type of representation, all the data are superimposed on the same master curve.}.
   \label{fig:spectro_master_curve}
\end{figure}
In Fig. \ref{fig:spectro_master_curve}, we have  plotted  these thicknesses as a function of the height $H(t)$ reached by the films during the generation, \textit{i.e.}  as a function of the product $V \times t$, for two different spectrometer positions. This leads to a reasonable collapse of all the data on a master curve.
The thickness evolution thus does not depend on the imposed velocity, which rules out inertial and air friction effects in the dynamics. The model will show that the central film  can indeed  be described by a hydrostatic equilibrium, which  can be disconcerting at first sight since we generate the films at high velocities. 

\subsection{Film thickness profiles} \label{part:thickness_profils}

\begin{figure}[htbp]
  \centering
    \includegraphics[width=0.7\linewidth]{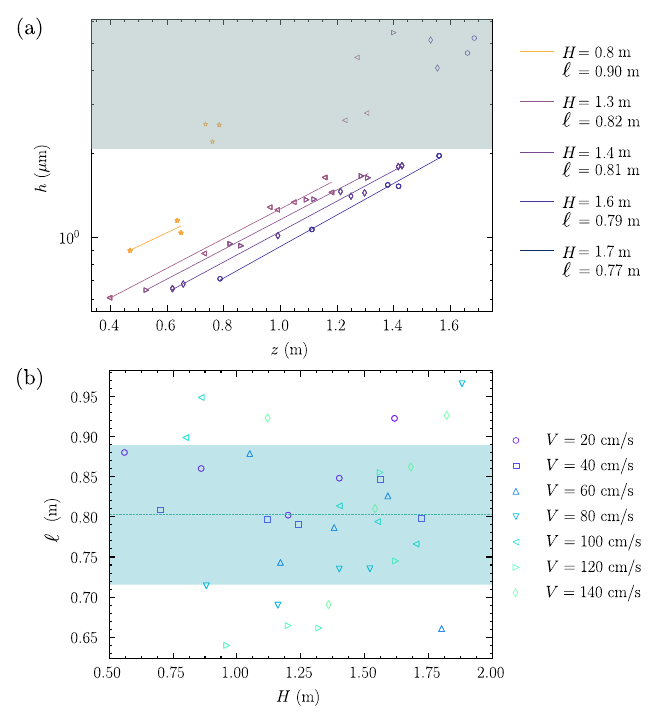}
 \caption{(a) Thickness $h$ of the soap films measured during the generation as a function of $z$ (oriented downwards, with the origin at the top of the film), in semi-logarithmic scale, for a velocity fixed at 100 cm/s. For a given instant, corresponding to a color, the value of $\ell$ extracted from the exponential fit and the height of the films $H$ are written in the legend. The average value is $\ell \simeq 0.82$~m. The points at the bottom of the films (3~cm from the bath) not described by the exponential are shown in the grayed area.
(b) Measurement of the characteristic length $\ell$ on the thickness profiles for different velocities (ranging from 20~cm/s to 140~cm/s). The values of $\ell$ are plotted as a function of the height reached by the soap $H$ films during this generation. The average value of $\ell$ measured for all the probed velocities is indicated by a blue line. The experimentally determined standard deviation is represented by the colored rectangle. All our measurements lead to $\ell = 0.80 \pm 0.09$~m. }
   \label{Fig_ExpProfils}
\end{figure}
We measured the thickness as a function of time for several positions of the spectrometer (between 3 and 180~cm above the liquid bath, in the center film in the $y$ direction) in order to determine the whole film profile, in the central part of the film. 
These profiles are given for five successive times, and therefore for five different film heights, in Fig. \ref{Fig_ExpProfils} (a). 
Two zones can be identified, despite a limited spatial resolution due to the difficulty of the thickness measurement in films moving at high velocity. At small $z$, close to the top of the film   (white background in Fig. \ref{Fig_ExpProfils} (a)), the profile can be fitted with an exponential model, whereas at large $z$, close to the liquid bath (gray background), the thickness is systematically larger than the extrapolation of this exponential profile.
In the following, the exponential part will be referred to as the initial film. Indeed, at short time, a small quantity of liquid is entrained by the fishing line and forms an initial film. As we will see later, this initial film is a closed system, which is stretched by the fishing line movement and by the film weight. The physical process governing this initial film extraction seems to differ from the visco-capillary law (Eq. \eqref{eq:loi_Frankel})  relevant at larger times.

The measurements made for seven entrained velocities, ranging from 20~cm/s  to 140~cm/s, lead to the same results. 
The top part of the profiles have been fitted by the law
\begin{equation} \label{eq:manip_exponentielle}
h (z) = K\,  \text{e}^{(z-H)/\ell} 
\end{equation}
with $\ell$ a characteristic length and $K$ a prefactor.  
These lengths are plotted in Fig. \ref{Fig_ExpProfils} (b) as a function of the film height (during the film generation). 
The values of $\ell$ obtained suggest that the characteristic length remains constant during the film's generation and does not depend on the entrained velocity.
Taking into account all the experiments, we measure \ER{an average} characteristic length $\ell = 0.80 \pm 0.09$~m. 

In the following (part \ref{Part:theoretical_description}) we will develop a hydrostatic theoretical description of the films during their generation to rationalize these observations.

\subsection{Area of lateral Frankel films} \label{Part:subs_aire_Frankel}

\begin{figure}[!ht]
  \centering
    \includegraphics[width=0.65\linewidth]{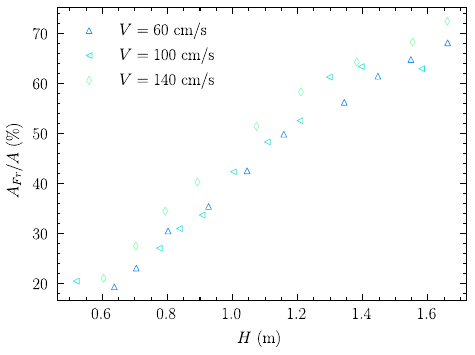}
 \caption{ Evolution of the proportion of the area of the lateral Frankel films $A_{\text{Fr}}$ to the total area of the films $A$ as a function of the height $H$ reached by the soap films during their generation for three different velocities: 60~cm/s, 100~cm/s and 140~cm/s. 
}
   \label{Fig_ExpAreas}
\end{figure}

We determined the area $A_{\text{Fr}}$ of the lateral Frankel's films by direct visualization, by dividing the films as described in the section \ref{part:direct_vizualisations}.
The ratio between $A_{\text{Fr}}$ ($A_{Fr}$ takes into account both sides of the film) and the total area  $A$ of the film  is given as a function of the film height $H$ for three different velocities in Fig. \ref{Fig_ExpAreas}. 
This ratio, like the central film thickness, is governed by the film height  and is independent of the entrained velocity.

\subsection{Surface tension gradients in the films}
\label{Part:Gene_apres_Grad_gamma}

As explained in the introduction, a vertical surface tension gradient due to gravity is expected along the giant soap film, which we propose to measure in this section. We perform this measurement just after the motors stop, when the film height reaches 120~cm.

The surface tension  is deduced from the deformation of a small elastic object, put  at different heights in the film. This method was initially proposed by Adami \textit{et al.} \citep{Adami2015}, and used in 15~cm high fed films. 

\begin{figure}[!ht]
  \centering
    \includegraphics[width=\linewidth]{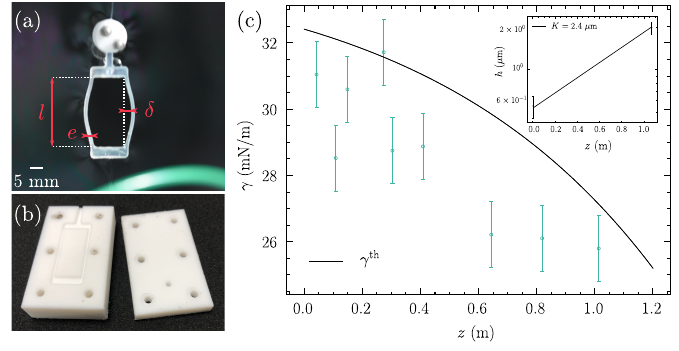}
  \caption{(a) Example of an image recorded during an experiment. The arms of the elastic object are deformed due to the surface tension forces. (b) Photograph of the PTFE mold used to fabricate the elastic object. (c) Surface tension $\gamma$ measured at the center (in the $y$ direction) of a film of height H=120~cm as a function of the coordinate $z$, just after the motors stop.  The black line corresponds to the surface tension $\gamma^{\text{th}}$, obtained by integrating Eq. \eqref{eq_films_gravite_vs_grad_Lucassen} using the film thickness profile given in inset, with the equilibrium tension at the film bottom. Inset: film thickness obtained by spectrometry at two different high. The whole profile (black line) is the prediction of Eq. \eqref{eq:manip_exponentielle}  with $\ell = 0.8$~m taken from Fig. \ref{Fig_ExpProfils} and $K = 2.4$~$\mu$m as adjustable parameter.
   }
   \label{Fig:mesure_gamma_avec_H}
\end{figure}

The elastic object used for the measure is shown in Fig. \ref{Fig:mesure_gamma_avec_H} (a). 
The sensor is made with Silicone SKIN FX 10 (from Rougier \& Plé), using a  PTFE mold (see  Fig. \ref{Fig:mesure_gamma_avec_H} (b)). This matter is usually used to make prosthesis or special effects in movies, and makes it possible to manufacture elastic objects in less than an hour.

The obtained sensor is held at the chosen height in the plane of the film by a thin wire attached to the horizontal fishing line entraining the films, and the piece of film trapped inside is pierced to perform the measure. The unbalanced tension forces outside the object deform the arms, as its shape tends to get closer to a circle. This deformation is quantified by the distance $\delta$ defined in Fig. \ref{Fig:mesure_gamma_avec_H} (a).
The deformation $\delta$ varies linearly with the tension around the object \citep{Adami2015}:
\begin{equation} \label{eq:Adami_delta_gamma}
\gamma \ = \ \frac{16  e^4  E}{l^4} \ \delta
\end{equation}
\noindent
with $e$ the thickness of the arms of the elastic object ($e = 0.8$~mm), $E$ the Young's modulus of the object ($E \simeq 0.4$~MPa) and $l$ the height of the arms ($l = 24$~mm). 
 As the Young modulus of the object is not precisely known, the prefactor of the linear relation \eqref{eq:Adami_delta_gamma} is obtained from a calibration  with horizontal soap films of known surface tension: a solution containing TTAB at a concentration of 10~cmc ($\gamma = 39$~mN/m), a solution containing Fairy diluted to 4 \% ($\gamma = 25.2$~mN/m) and a solution containing Tween 20 at 1 \% ($\gamma = 33.7$~mN/m). 

From an experimental point of view, this measurement is a real challenge because the films last only a few seconds. 

The results obtained for a 120~cm high film, entrained at a velocity of 100~cm/s, are shown in Fig. \ref{Fig:mesure_gamma_avec_H} (c). The measurements begin just after the motors stop, and last 10 seconds. The thicknesses given in the inset are average during this time period. We measure a surface tension difference, between the top and bottom of the film, of about 5~mN/m, which we will compare to a theoretical prediction in the next paragraph.

\section{Theoretical description of the soap films during their generation} \label{Part:theoretical_description}

\subsection{Equilibrium film thickness profile under gravity}
\label{sec:eq}

A static model build by Lucassen \citep{Lucassen1981} and used by de Gennes \citep{Gennes2001} relies on the fact that the weight of the soap films is balanced by the surface tension gradient so that
\begin{equation} \label{eq_films_gravite_vs_grad_Lucassen}
\rho g h  + 2 \ \frac{\partial \gamma}{\partial z} = 0 \, , 
\end{equation}
\noindent with $h$ the thickness of the soap film and with the vertical axis $z$ oriented downwards.

This equation is simply a force balance in the vertical direction, per unit area, on a film element of volume $\de V = h \, \de A$ spanning from one interface to the other. 

The validity of Eq. \eqref{eq_films_gravite_vs_grad_Lucassen}  can be tested using 
the tension measurements made in 120~cm high film, described in the section \ref{Part:Gene_apres_Grad_gamma}. 
To obtain the theoretical profile shown in Fig. \ref{Fig:mesure_gamma_avec_H} (c), inset, we assumed an exponential variation of the thickness with $z$. The measurements are compatible with the characteristic length $\ell = 0.8$~m,  as obtained for all films during generation (see Fig. \ref{Fig_ExpProfils}). 
The theoretical tension profile is then obtained by integration of Eq. \eqref{eq_films_gravite_vs_grad_Lucassen}, using the theoretical profile and $\gamma_0 = 25.2$~mN/m, $\rho = 997$~kg/m$^3$ and $g = 9.81$~m/s$^2$. 
The measure, shown in Fig. \ref{Fig:mesure_gamma_avec_H} (c), is slightly below the prediction proposed, but with an agreement close to the experimental uncertainty.

The surface tension involved in Eq. \eqref {eq_films_gravite_vs_grad_Lucassen} depends on the interface surfactant concentration $\Gamma$.  For small coverage variations, an interface elasticity $E$ is defined by Eq. \eqref{Eq_def_EGibbs} and a linearized expression of the surface tension can be expressed as  $\gamma = \gamma_0 - E (\Gamma-\Gamma_0)/\Gamma_0$.

Here, we will focus on films stabilized by a dishwashing liquid and we will assume that the film elasticity is mainly due to poorly soluble surfactant molecules, of negligible bulk concentration. 
In that case, a simple relationship between $E$ and the film mechanical dilatationnal elasticity, {\it i.e.} between the film tension and dilatation or compression can be obtained, once the reference state of each  film element is defined.

In this aim, we consider a film element of volume $\de V$ and define its reference state, indicated by the subscript 0, as the state in which the interface concentration is in equilibrium with the bulk liquid used to produce
the film. It is characterized by its area $\de A_0$, its  thickness $h_0$ and the interface concentration $\Gamma_0$. 
The relative motion of the bulk phase with respect to the interfaces is
a downwards Poiseuille flow, with a velocity scaling as $\rho g h^2/\eta \sim 10 \ \mu$m/s,  which is entirely negligible in our experiment. A film element is thus a closed material system.
Consequently, when the film element is stretched to reach an area $\de A$, the volume conservation imposes $h \de A = h_0 \de A_0$. Moreover,  the surfactant conservation leads, \ER{considering insoluble surfactants}, to $h \Gamma = h_0 \Gamma_0$ and finally

\begin{equation} \label{Eq_relation-surface-tension_and_extension_thickness}
\gamma = \gamma_0 \ + \ E \frac{\varepsilon}{1 + \varepsilon}
 \end{equation}
with the  extension $\varepsilon$  defined by :
\begin{equation} \label{Eq_def_epsilon}
\varepsilon = \frac{\de A}{\de A_0} \ - \ 1 = \frac{h_0}{h} \ - \ 1 \, .
\end{equation}
As there are two interfaces, the film tension is simply $2 \gamma$.

From these relations, we get $\partial_z  \gamma = - E \partial_z (h/h_0)$, and  Eq. \ref{eq_films_gravite_vs_grad_Lucassen}  becomes  
%
\begin{equation} 
\label{eq:h}
\dfrac{1}{h} \dfrac{\partial }{\partial z} \left ( \frac{h}{h_0} \right)  \ = \dfrac{ \rho g}{2 E} \, . 
\end{equation}
In the case of soluble surface-active molecules, the film elasticity also depends on the film thickness \cite{Prins1967, tempel_application_1965, Lucassen1981}. In the frame of this model, the thickness profile of the soap film is a 2D equivalent of a density profile in a 3D compressible gas in a gravity field. 

The Eq. \eqref{eq:h} governs the thickness profile of a film in a gravity field. 
It predicts an exponential profile only if the reference thickness $h_0$ does not vary with the position $z$. Indeed, in this case it can be integrated to get  
\begin{equation} \label{Eq_h(z)_analytic_sol_with_unknown}
h (z) = h(H)\,  \text{e}^{(z-H)/\ell} \quad \text{with} \quad \ell = \frac{2 E}{\rho g h_0}  
\end{equation}
the characteristic length of the problem. The exponential profile observed in Fig. \ref{Fig_ExpProfils} is thus the signature of a homogeneous reference thickness in the initial film. 

This reference thickness is governed by the amount of surfactant trapped is the film during its extraction. As this reference thickness is uniform in the upper part of the film (defined as the initial film), the extraction process should thus remain identical during the production of the whole initial film. 
Even more surprisingly, the reference thickness, and thus also the initial extraction process, does not depend on the motors' velocity. 
This is an important experimental result that we have not been able to elucidate yet, as discuss in the conclusion.

Moreover, as verified in Fig. \ref{Fig_ExpProfils} (b), the characteristic length $\ell$ is constant with time. This is in contrast an expected result, consistent with the fact that a film element is a closed material system, keeping the same reference thickness at all times. 

Once the reference thickness is known, the profile of a film is entirely determined by a boundray condition, for example the thickness at the bottom of the film $h(H)$ (in the case of Eq. \eqref{Eq_h(z)_analytic_sol_with_unknown}).
We will show that it only depends on the area and on the shape of the film.

\subsection{Validity of the equilibrium force balance} \label{part:eq_balance_validity} 

The Eq. \eqref{eq_films_gravite_vs_grad_Lucassen} is the projection in the vertical direction of the vectorial equilibrium condition of the film  $\rho g h \vec{e}_z + 2\vec{\nabla} \gamma =0$. Taking the curl of this equation, we obtain $\partial_y \gamma = 0$, with $y$ the horizontal direction in the film plane. Now using $\rho g h = - 2 \partial_z \gamma$ we obtain, after a derivation, the condition $\partial_y h=0$. This condition is verified in the central part of the film, but not  in the lateral part, which  is thus necessary in an out of equilibrium situation (see Fig. \ref{Fig:photos_films-geants_montrant-zones-laterales}).

The air friction on a moving film can be estimated using the Prandtl theory  \cite{rutgers96}. 
It scales as $\eta_a V / \delta_{\text{BL}}$ with the boundary layer thickness $ \delta_{\text{BL}} \sim [\eta_a H /(\rho_a V) ]^{1/2}$, the air viscosity $\eta_a = 1.8 $ kg/m/s, the air density $\rho_a = 1.3$ kg/m$^3$, and the film height $H \sim 1$ m. 
In the initial film, the velocity varies from the motors' velocity at the top to a much smaller velocity at the bottom (as will be shown later). For the smallest motors' velocity, $V= 0.2$ m/s, the air friction is thus at most of the order of 10$^{-3}$ kg m$^{-1}$ s$^{-2}$ and should be negligible in comparison with the gravity of the order  $\rho  g  h \sim 10^{-2}$ kg m$^{-1}$ s$^{-2}$. 
The highest motor velocities are close to the terminal velocities observed for films of comparable thickness (see Fig. 4 in \cite{rutgers96}) and air friction should begin to play a significant role. However, the rescaling obtained in Fig. \ref{fig:spectro_master_curve} shows that the influence of the motors' velocity is small, and the air friction will be neglected in the model.
The inertia of the central film scales as $ \rho h V^2 /H \sim 10^{-3}$ kg m$^{-1}$ s$^{-2}$ and will be neglected too.

Finally, the surface shear viscosity leads to a vertical in-plane friction force of the order of $ \eta_s \delta V_z / L^2$, with $\delta V_z$ a velocity difference between two points of the film separated from a horizontal distance $L$ and $\eta_s \sim 10^{-7} - 10^{-6}$ kg/s \cite{Stevenson2005} the shear interface viscosity. The thickness stratification in the central part of the film leads to an invariance by translation in the $y$ direction, and this viscous friction does not contribute to the force balance. The situation is different in the   extracted films on the two film sides that are thicker than the central part and fall down under gravity.
If the downward velocity $V^{\text{Fr}}$ reaches 1 m/s, with a characteristic width \ER{ $L \sim 10^{-1}$ m (width of the lateral Frankel films)} we get an upward viscous term of the order of $10^{-4}$ kg  m$^{-1}$ s$^{-2}$ in the Frankel film, which is negligible. The higher gravitational force in this thick part of the film is thus probably compensated by the film inertia.

On the basis of the previous discussion and of the observed rescaling in Fig. \ref{fig:spectro_master_curve}, the dynamics of the initial film will be rationalized using the equilibrium model of section \ref{sec:eq}. The modeling of the Frankel film motion is beyond the scope of this paper, and its downwards flow will be taken into account phenomenologically.  

\subsection{Modeling of the time evolution} \label{part:model_extraction}

The goal of this section is to predict the time evolution of the film thickness profile, from the end of the non-quasistatic regime to the end of the motors' motion, on the basis of the quasistatic model leading to Eq. \eqref{eq:h}. The profile at the initial time is fitted on the experimental results, and its subsequent dynamics is predicted from an interplay between gravity, motors' motion and film extraction. 


\subsubsection{Film parametrization and equilibrium equation} 

\begin{figure}[!ht]
  \centering
  \includegraphics[width=0.9\linewidth]{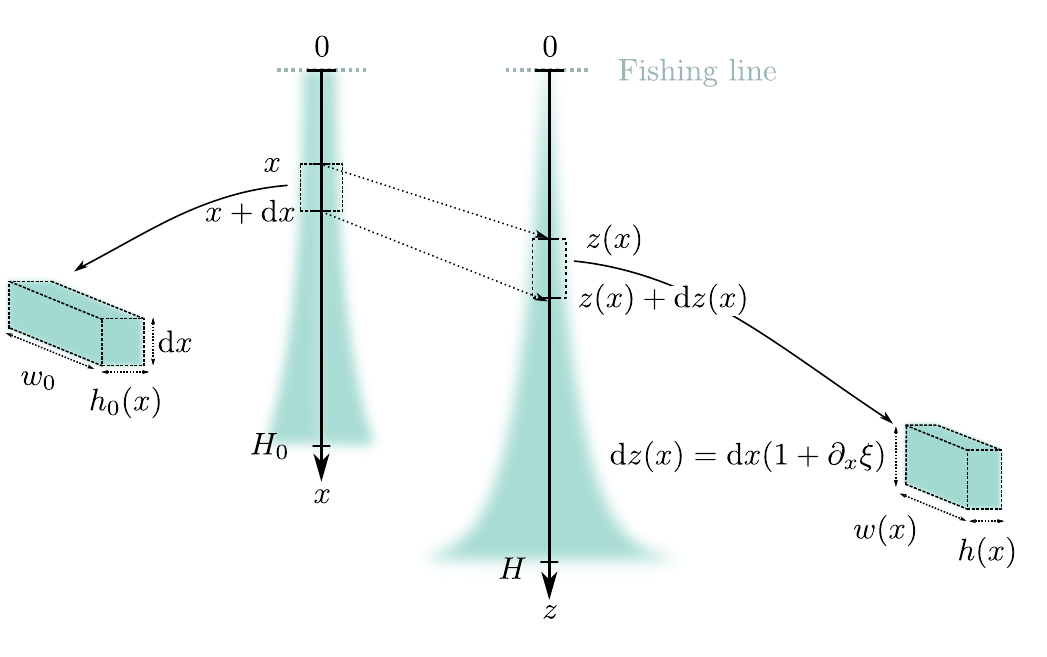}
\caption{Scheme of the film profile and notations used in the text. (Left) Reference state of the film with a thickness $h_0 (x)$ and a height $H_0$. An elementary  material system  is defined between $x$ and $x+ dx$.  (Right) Actual state of the film, in which the same elementary system is located between $z(x)$ and $z(x)+ dz(x)$.
}
   \label{Fig_shema_notations_modele_complete_films-geants}
\end{figure}

The central film is defined as the material system extracted from the bottom meniscus  stretched by the fishing line entrainment. We assume that it remains invariant by translation in the horizontal direction, so the relevant elementary systems are   horizontal film strips of homogeneous properties.
These film elements are identified by their distance $x$ from the top meniscus, measured in the reference state of the film, using a Lagrangian approach. 
$x$ is the position the film element would have in the absence of gravity and with a film tension $2 \gamma_0$ imposed at its boundaries (see Fig. \ref{Fig_shema_notations_modele_complete_films-geants}). With this definition the film element of height $\de x$ is a closed material system, that is followed during its motion and deformation. This Lagrangian approach leads us to introduce the spatial variable $\xi$ defined such that the position $z$ of the tracked element in the film is related to the variable $x$ \textit{via} the relation:
\begin{equation}
z (x) = x + \xi(x)
\end{equation}

The shape of the film element is a rectangle $\de x (1+ \partial_x \xi) w(x)$, with $w(x,t)$ the width of the central film, from one boundary with the lateral extracted film to the other. 
The reference thickness $h_0(x)$ of the element of reference position $x$ is determined by its extraction from the bath, which occurs at the time $t^{\text{ex}}(x)$. On the basis of our experimental results, it will be assumed uniform in the top part of the film, and then governed by the visco-capillary film extraction (Eq. \eqref{eq:loi_Frankel}). Note that the Lagrangian variable $x$ corresponds to a material system which keeps a reference thickness constant with time, so $h_0$ may depend on $x$ but not on time. The width of the element at $t^{\text{ex}}(x)$ is $w_0$, as the film extraction  occurs along the whole  bottom meniscus length. This length is chosen as the reference width for the film element.

The volume $\de V$ of a film element is conserved during a deformation, which allows us to write
\begin{equation}\label{vol_cons}
w_0 \ h_0(x) = ( 1 + \partial_x \xi ) \ w(x) \ h(x).
\end{equation}

Its area $\de A= w(x)( 1 + \partial_x \xi ) \de x$   may in contrast vary and the film extension $\epsilon (x)$ (Eq. \eqref{Eq_def_epsilon}) is:
\begin{equation} \label{Eq_def-epsilon(x)_en_fct_params}
\epsilon (x) = \dfrac{w(x) \ ( \ 1 + \partial_x \xi )}{w_0} \ - \ 1.
\end{equation}

Within the central part of the film, the force balance is given  by Eq. \eqref{eq_films_gravite_vs_grad_Lucassen} and \eqref{Eq_relation-surface-tension_and_extension_thickness} which becomes: 

\begin{equation} \label{eq:balance14}
\rho g h (x) \ ( 1 + \partial_x \xi ) \ = \ - 2 \ E \ \partial_x \left( \frac{\epsilon (x)}{ 1 + \epsilon (x) } \right)    \, .
\end{equation}
Using Eq. \eqref{vol_cons} we can reformulate this equation as a function of $h_0$, $w$ and $\varepsilon$ only 

\begin{equation}\label{equili}
 \partial_x \left( \frac{\epsilon (x)}{ 1 + \epsilon (x) } \right) = - \frac{1}{\ell} \frac{h_0(x)}{h_{00}} \frac{w_0}{w(x,t)}  \, ,
\end{equation}
\ER{The Eq. \eqref{eq:balance14} contains a single parameter $ 2E / (\rho g) $. In order to built the  relevant length ratio in Eq. \eqref{equili}, we use $\ell$  the characteristic length obtained experimentally in the exponential part of the profiles (see section \ref{sec:eq}), and define $h_{00}$ as $ h_{00} = 2E / (\rho g \ell)$.
With this definition $h_{00}$ is the uniform reference thickness of the exponential part of the film (see Eq. \eqref{Eq_h(z)_analytic_sol_with_unknown}).  
This arbitrary decomposition does not reduce the generality of  the problem, as the two length scales only appear as their product $\ell h_{00}$. It provides an explicit expression for the relevant length scale of the problem. }

The Eq. \eqref{equili} determines the extension $\varepsilon$ and thus the whole film profile if the function $w(x,t)$ and $h_0(x)$ are known, as well as a boundary condition at the film bottom. 
These quantities are determined below, from the film extraction at the bottom and lateral menisci, and from the variation of the film height $H$ imposed by the motors.

\subsubsection{Film extraction} 

 In the limit of high  film elastic modulus, and in a steady state regime, a film extraction at a velocity $U_{Fr}$ is associated with a film tension $2 \gamma_0 + 2 \Delta \gamma$, with $\Delta  \gamma$ given by Eq. \eqref{eq:Frankel_deltagamma}.

The extraction rate can thus be related to the local tension in the film, and using  Eq. \eqref{Eq_relation-surface-tension_and_extension_thickness}, to its deformation:

\begin{equation} \label{eq:U_Fr_meniscus_low_simus}
U_{\text{Fr}} (x) = \beta \left( \frac{\epsilon(x)}{1 \ + \epsilon(x)}\right) ^{3/2} \quad \text{with} \quad \beta \ = \ k \ \frac{E^{3/2}}{\eta \gamma_0^{1/2}}
\end{equation}
\noindent
and $k$  a numerical prefactor which theoretical value, deduced from Eq. \eqref{eq:Frankel_deltagamma}, is $k=0.13$. 

The above equation assumes that the surface tension in the menisci is everywhere equal to the reference surface tension $\gamma_0$. This is a strong assumption, especially for the lateral menisci. If these menisci are not  sufficiently effective surfactant reservoirs and become depleted, there can be a large error in the extraction velocity of the side films. To overcome this difficulty, we have chosen to use \ER{$\beta$} as a fitting parameter. For the physical situation studied in this article, the theoretical prediction of $\beta$ with the Eq. \eqref{eq:U_Fr_meniscus_low_simus} is: $\beta = 2. 63$~m/s, with $E = 22$~mN/m \cite{Kim2017}, whereas the fitted value will be $\beta = 0.25 \pm 0.05$~m/s.  
Note that this expression is only valid for a positive extraction velocity, so for a stretched film, at a tension higher than its surrounding meniscus.

During the dynamics, some new film elements are added in the system, thus increasing the range value for the Lagrangian coordinate $x$ which varies between $x=0$ at the top and $x=H_0(t)$ at the bottom. 
\ER{The extraction rate $U_{\text{Fr}} (H_0)$} is associated to the increase of $H_0$ at the rate 
\begin{equation} \label{Eq_dx-Frankel_extract_bain}
\frac{\de H_0}{\de t} = \frac{U_{\text{Fr}}(H_0)}{1 + \epsilon (H_0(t)) }
\end{equation}

The thickness $h_{\text{Fr}}$ of the extracted film elements is governed by Eq. \eqref{eq:Frankel_deltagamma} and  thus  depends on the local deformation $\epsilon (H_0)$ through Eq. \eqref{eq:U_Fr_meniscus_low_simus}. The reference thickness $h_0$  can then be computed at $x= H_0(t)$ using Eq. \eqref{Eq_def_epsilon}   

\begin{equation}\label{h0H0}
h_0(H_0) = h_{\text{Fr}}(U_{Fr}(H_0)) \ \times \ ( 1 + \epsilon (H_0))
\end{equation}

The mechanism at the origin of film extraction from the lateral meniscus is the same as the one at the origin of the extraction from the bottom meniscus.

The width of the lateral extracted film $w_{Fr}$ thus increases at the rate  $\partial_t w_{Fr} (x) =    U_{\text{Fr}} (x)$ 
and its area  varies as
\begin{equation}\label{dAFr}
\frac{\de A_{Fr}}{\de t} = \int_0^{H_0}  U_{\text{Fr}} (x) (1 + \partial_x \xi) dx
\end{equation}

With this model the lateral film width $w_{Fr} (x)$ should be an increasing function of $x$, first because the extraction begins sooner at smaller $x$ (because the film element is created earlier), and second because the tension, and thus the extraction velocity, is higher at the top of the film. 

We observe experimentally that these lateral films are actually wider in the central part (see Fig. \ref{Fig:photos_films-geants_montrant-zones-laterales}). This is due to the fact that they are thicker than the central part of the film, which leads to their fast downward motion. 
As this dynamical effect is beyond our theoretical description, we choose to take it into account through a phenomenological correction of the model. The area evolution of the lateral extracted film  is still the one predicted by Eq. \eqref{dAFr}, 
 but its shape  is assumed  to be a rectangle of uniform  width $\bar{w}_{Fr}$ verifying $H \bar{w}_{Fr} = A_{Fr}$. This phenomenological redistribution of the lateral film elements mimics the downward lateral motion in the simplest possible way, and keeps a realistic shape for the central part of the film, which is the system of interest. The central film  width is thus assumed to be uniform, with $w(t)= w_0 - 2  \bar{w}_{Fr}$.

 \subsection{Numerical resolution} \label{Part_Modele-geant-films-numerical-resolution}
 \subsubsection{Equations set}
 
 The film profile is determined by its reference thickness $h_0(x)$ and by its  deformation $\epsilon(x,t)$. 
As the model is based on an equilibrium equation, it only evolves because of the boundary conditions evolution: the variation of its width   $w(t)$ and of its reference height $H_0(t)$.
At each time, $h_0$ is obtained at the bottom of the film from Eq. \eqref{h0H0} and $w(t)$ is deduced  from Eq. \eqref{dAFr}. 
The Eq. \eqref{equili} can thus be numerically integrated if the tension is known at a given point of the film.

The missing information is obtained from the total height of the film, imposed by the motors motion: 
\begin{equation} \label{Eq_relationH-and-xi}
H(t) =  \int_0^{H_0}(1+ \partial_x \xi(x)) \de x
\end{equation}
By using Eq. \eqref{Eq_def-epsilon(x)_en_fct_params}, we can express $\partial_x \xi(x)$ as a function of $\epsilon (x)$, leading to 

\begin{equation} \label{Eq_relationH-and-eps}
H(t) =  \int_0^{H_0}\frac{w_0}{w(x)} (1+ \epsilon(x)) \de x
\end{equation}
Introducing the auxiliary function 
\begin{equation} \label{Eq_def_f0-and-f(x)}
 f(x) = \frac{\epsilon(H_0)}{1 + \epsilon(H_0)} - \frac{\epsilon(x)}{1 + \epsilon(x)}
\end{equation}
we get, from Eq. \eqref{equili}, 
\begin{equation} \label{Eq_modele_geants_eq-on-f}
f(x) = - \frac{1}{\ell} \int_x^{H_0}\frac{h_0(x)}{h_{00}} \frac{w_0}{w(t)} \de x
\end{equation}
The function $f$ is thus expressed as a function of the known quantities $h_0$ and $w$.
The Eq. \eqref{Eq_relationH-and-eps}  leads to 
\begin{equation} \label{Eq_modele_films_geants_eq-on-H}
H(t) = \int_0^{H_0} \frac{w_0}{w(t)} \frac{1}{1 - f_0 + f(x)} \ \de x ,
\end{equation}
which provides an implicit equation for the parameter $f_0 = \epsilon(H_0)/(1 + \epsilon(H_0))$. Once $f_0$ has been determined from Eq. \eqref{Eq_modele_films_geants_eq-on-H}, the film extension can be deduced from Eq. \eqref{Eq_def_f0-and-f(x)}, which closes the model.

The problem can in principle be solved from an initial reference film height $H_0(0)=0$, thus addressing the first moments of the film extraction from the bottom meniscus. However, using our experimental parameters, this model leads to unrealistic prediction for the first 10 cm. \ER{Indeed, an extraction rate of 1~m/s corresponds, using Frankel's model (Eq. \eqref{eq:loi_Frankel}), to a thickness of 350 $\mu$m, which is much larger than the thicknesses measured experimentally in soap films. We assume that inertial effects are non-negligible at this stage, making our quasi-static model irrelevant.}
We thus initialize the code with a small film of height $H_0$, and of uniform thickness $h_0$ and width $w_0$.
\ER{These parameters are adjustable parameters deduced from the experimental results.}
 This initial film corresponds to the liquid entrained by inertia and viscosity at very short time.

 \subsubsection{Numerical scheme}

The  numerical resolution of the model was carried out under Python. 
The film is discretized in $n$ strips. Their reference height $\de x$ is uniform, except for the bottom strip which has a variable height  $\de x^{\ast}$. 
At each time step, the height $H$ is increased by $V \de t$, with $V$ the  velocity of the motors. We compute $f(x)$ in each film element \textit{via} Eq. \eqref{Eq_modele_geants_eq-on-f}. We determine the tension at the bottom meniscus $\epsilon(H_0)$ (Eq. \eqref{Eq_def_f0-and-f(x)}) by solving Eq. \eqref{Eq_modele_films_geants_eq-on-H} to find the value of $f_0$. The tension $\epsilon(x)$ in each element is deduced from $f(x)$ using Eq. \eqref{Eq_def_f0-and-f(x)}. The thickness $h(x)$ is computed using Eq. \eqref{Eq_def_epsilon}. If the tension at the bottom of the film is larger than $\gamma_0$,  the film length  $\de z_{\text{Fr}} = U_{\text{Fr}} \ \de t$ is extracted from the bottom  meniscus.  The height of the last element $\de x^{\ast}$ is thus increased by 
\begin{equation} \label{Eq_dx-Frankel_extract_bain}
\de x_{\text{Fr}} =  \frac{U_{\text{Fr}} \ \de t}{1 + \epsilon (H_0) } \, .
\end{equation}
and the reference thickness $h _0(H_0)$ of the last segment is recalculated  so that $h(H_0) = h_{\text{Fr}}$ (Eq. \eqref{eq:loi_Frankel}) and $h (H_0) = h _0(H_0) / (1+ \epsilon(H_0))$. 
As soon as $\de x^{\ast} \geq 2 \de x$, the strip $n$ is divided in two: the strip $n$ with a reference height $\de x$ and the strip $n+1$ with the reference height $\de x^{\ast} \ - \de x$. 
The height of the film at rest  is recalculated as $H_0 = n \de x$.
The increase of the lateral extracted film is computed from  Eq. \eqref{dAFr}. 
The central film width is then obtained from 
\begin{equation}
w = w_0 - 2 \ \frac{A_{\text{Fr}}}{H}.
\end{equation}

\begin{figure}[!ht]
  \centering
    \includegraphics[width=0.65\linewidth]{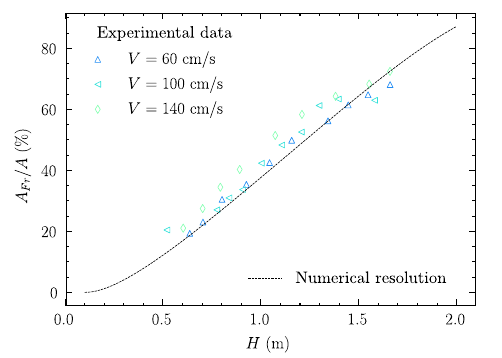}
 \caption{Evolution of the proportion of the area of the lateral Frankel films $A_{\text{Fr}}$ to the total area of the films $A$ as a function of the height $H$ reached by the soap films during their generation for three different velocities: 60~cm/s, 100~cm/s and 140~cm/s (data identical to those shown in Fig. \ref{Fig_ExpAreas}). The black dashed line is the numerical resolution of the model for $V = 100$~cm/s and $\beta = 0.26$~m/s. }
   \label{Fig_ExpAreasvsNum}
\end{figure}

\subsection{Results of the numerical resolution}  \label{Part:resolution_numerique}

\subsubsection{Choice of the numerical parameters}

There are two numerical parameters in the equations set: the ratio $2E/(\rho g) = h_{00} \ell$ (Eq. \eqref{eq:balance14}, \eqref{equili}), and the prefactor $\beta$ (Eq. \eqref{eq:U_Fr_meniscus_low_simus}). Their values, as well as the film reference height and thickness at short time, have been determined from the experimental data, as discussed below.

\ER{Using $E= 22$~mN/m \cite{Kim2017}, we get $2E/\rho g = 4.4 \times 10^{-6}$ m$^2$. 
In order to obtain an exponential profile of characteristic length $\ell = 0.8$~m in the top part of the film (Fig. \ref{Fig_ExpProfils}), we need to impose a reference thickness $h_0 = h_{00} =  2E/(\rho g \ell) = 5.6 \ \mu$m (see Eq. \eqref{Eq_h(z)_analytic_sol_with_unknown}). 
The size of the exponential part of the film is recovered by choosing an initial film reference height $H_0 = 10$ cm.}

The parameter $\beta$ is determined from the experimental data shown in Fig. \ref{Fig_ExpAreasvsNum}: the proportion of the Frankel film area as a function of time obtained numerically fits the data for $\beta = 0.26$~m/s. This value is ten times lower than the one expected if the surface tension in the menisci was equal to that of the bath, which indicates that the lateral menisci are at a tension much higher than the equilibrium tension and that they are depleted in surfactants.

Finally, we choose to discuss the case $V= 100$~cm/s.

%

\subsubsection{Numerical results}

\begin{figure}[!ht]
  \centering
    \includegraphics[width=0.65\linewidth]{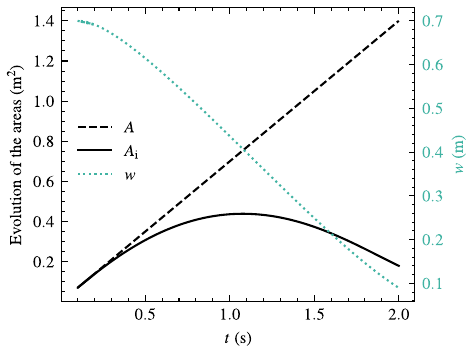}
 \caption{(Left) Evolution of the areas of the initial film $A_{\text i}$ (solid black line) and of the total film area $A$ (dashed black line) and (Right) evolution of the width $w$ (dotted green line) as a function of time, obtained numerically. }
   \label{XX}
\end{figure}

The numerical results allow us to quantify the various processes involved in the dynamics. We will first discuss the properties of the  initial film, which can be easily followed at different times, using our Lagrangian formalism.  A salient feature is the competition between the film extension, due to the motors' motion, and the film extraction, which mitigates this extension.
 This competition is clearly illustrated by the initial film area, plotted in Fig. \ref{XX} as a function of time:
it increases during the first second, remaining close to the total film area imposed by the motors' motion, and decreases at longer time, as the  Frankel films  are invading the films. The largest part of the extraction occurs in the lateral menisci, and the total area of the central film remains close to the area of the initial film. 

\begin{figure} [!ht]
    \centering
    \includegraphics[width=1\linewidth]{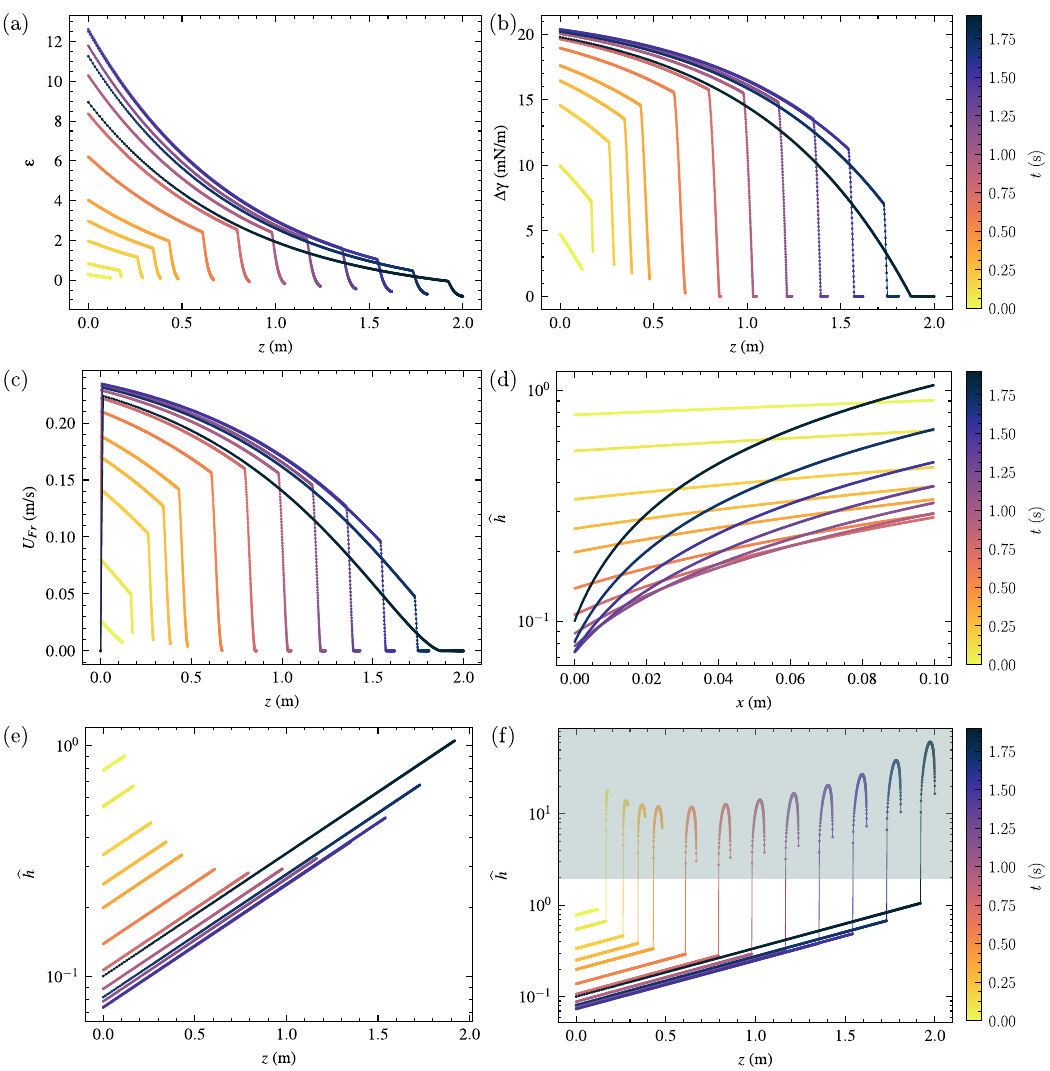}
    \caption[]{Numerical resolution results for the generation of a film of a width $w_0 = 70$~cm and an initial height for the reference state $H_0 = 10$~cm, at a velocity $V = 100$~cm/s. Each color corresponds to a different time $t$, indicated by the vertical color bar. The difference between two represented instants is $\Delta t = 0.19$~s. (\textit{see next page}) }
\end{figure}
\begin{figure} [!ht]
    \captionsetup{labelformat=adja-page}
    \ContinuedFloat
    \caption[Figure]{Evolution of (a) the extension $\epsilon$ as a function of the $z$ coordinate, (b) the tension $\Delta \gamma = \gamma - \gamma_0$ as a function of the $z$ coordinate, (c) the extraction velocity of the lateral Frankel films $U_{\text{Fr}}$ as a function of the $z$ coordinate, (d) the dimensionless thickness $\widehat{h} = h /h_{00}$ for the initial film as a function of the $x$ coordinate, (e) the dimensionless thickness $\widehat{h}$ for the initial film as a function of the $z$ coordinate, (f) the dimensionless thickness $\widehat{h}$ as a function of the $z$ coordinate for the whole film: the initial stretched film (whose data are represented in the figure (e)) and the Frankel's films extracted from the meniscus with the bath. The circles correspond to the points measured in each element. These resolutions are conducted by dividing the initial film into $n =$ 500 elements and setting the time step $\de t = 10^{-4}$~s. }
    \label{Fig:results_resolution}
\end{figure}

This non-monotonic evolution is also visible in Fig. \ref{Fig:results_resolution} (a), where the film extension is plotted as a function of $z$ for different times. The slope discontinuity in the curves corresponds to the boundary between the initial film and the film extracted from the bottom meniscus. 

In the initial film, the average extension is related to the total area of the initial film, whereas the extension gradient is due to gravity. The film tension shown in Fig. \ref{Fig:results_resolution} (b) is directly deduced from the film extension, using Eq. \eqref{Eq_relation-surface-tension_and_extension_thickness}, and governs the film extraction. Its value at the different heights in the film is a complex interplay between (i) the non-monotonic evolution of the mean tension, due to the motors motion and the film extraction,  and (ii) the increase of the tension difference between the top and the bottom of the initial film due to the gravity and to the increase of the initial film height. 
The resulting tension profile leads to the extraction rate from the lateral menisci plotted in Fig. \ref{Fig:results_resolution} (c): the higher we are in the film, the higher the tension and thus the higher the extraction velocity. 
It can be noted that, as we will be discussed later, we have chosen to set a zero extraction velocity at the bottom when the extension values are negative, and thus prevent the film from entering the bath meniscus. This explains the plateaus observed at long times in Fig. \ref{Fig:results_resolution} (b,c). 

The thickness profile of the initial film is the most important prediction, as it can be compared to the experimental data. Its dimensionless value $\widehat{h} = h /h_{00}$ is plotted using two representations in Fig.  \ref{Fig:results_resolution} (d,e). In Fig.  \ref{Fig:results_resolution} (d), we use the Lagrangian parameter $x$ so that a given piece of film can be followed during its motion and deformation. At each place in the initial film the non-monotonic behavior is observed, with a thinning followed by a thickening, but the transition occurs at a longer time at the top of the film. The profile is plotted as a function of the actual position in the film $z$, in Fig.  \ref{Fig:results_resolution} (e), and the exponential profile is recovered. This profile is a direct consequence of the initial condition we used: as a uniform reference thickness is imposed in the initial film, the profile remains exponential at all times. The prefactor is, however, an important prediction of the model, and shows again a non-monotonic evolution.  

All the results discussed above help describing the physics of the initial film evolution during the motors motion. 
In Fig. \ref{Fig:results_resolution} (f), we have represented the thickness profile of the whole central film (the initial film and the film extracted at the bottom at longer times).
The transition leads to a non-physical discontinuous film thickness, due to the fact that the initial film thickness and the extracted film thickness are governed by independent assumptions. As the tension gradient is governed by the film thickness, the thickness discontinuity leads to the slope discontinuity observed in Fig. \ref{Fig:results_resolution} (a) (extension). 
Finally, the model leads to compression at the bottom of the film, and to a tension lower than the equilibrium one. This corresponds to a tendency of the system to extract film at the top at high tension, and to absorb some film at the bottom meniscus, where the film tension is lower. Such film absorption is difficult to evidence experimentally, and could be associated with the marginal regeneration observed at the bottom of vertical soap films \citep{gros2021marginal}. However, the extraction velocity has been set to zero in that case in order to keep the model simple.

\section{Comparison with experimental data}  \label{part:comparaisonExpManip}

We propose in this part to compare the results of the numerical resolution presented in the previous section with the experimental results presented in the section \ref{part:thickness_profils}.
\begin{figure}[!ht]
  \centering
    \includegraphics[width=0.65\linewidth]{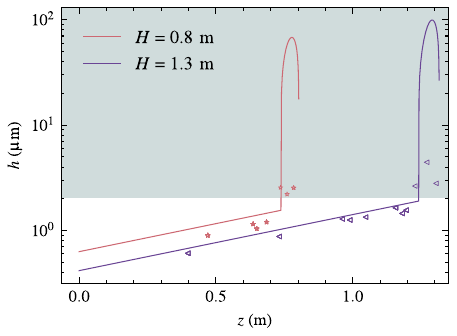}
 \caption{Evolution of the thickness $h$ as a function of $z$, in semi-logarithmic scale, for a entrained velocity fixed at 100 cm/s and two different heights: $H = 0.8$~m (pink) and $H = 1.3$~m (purple). The solid line corresponds to the numerical resolution while the points (triangles or stars) are experimental data identical to those in Fig. \ref{Fig_ExpProfils}.}
   \label{Fig_ExpNumProfils}
\end{figure}

In Fig. \ref{Fig_ExpNumProfils}, we have plotted two thickness profiles obtained numerically and experimentally for a film generated at a velocity fixed at 100~cm/s. A good agreement is obtained for the initial film profile. Note that the characteristic length has been fitted on the experimental data, but the prefactor of the exponential profile is quantitatively predicted, which is an important validation of the model.
 \ER{The thickness jump in the profiles corresponds to the boundary between the initial film and the thick Frankel film extracted at the bottom.}
In the film extracted at the bottom, the thickness obtained numerically is more than 10 times larger than the experimental one, which is a failure of the model.   This may be due to the  fact that we did not take into account the Poiseuille flow induced by gravity. Indeed, the downwards Poiseuille flow scales as $\rho g h^2/\eta $. Thus, for thicknesses between 10 and 100 $\mu$m gravity become non-negligible in the extracted film ($\rho g h^2/\eta \sim 0.1 - 10 \,  \ $cm/s). This explains that the measured thicknesses are smaller than the prediction. However, the width of this film is reasonably captured.

Importantly, the dynamics of the initial film, on which this paper focuses, is not affected by this non-physical film thickness at the bottom. The tension profile in the initial film is governed by the gravity and by the constraint that the initial film must reach a total height, determined by the top wire position and by the location of the boundary with the extracted film at the bottom. Only the size of the bottom extracted film matters for the initial film dynamics, not its thickness. As the bottom extracted film has a negligible size, both experimentally and numerically, the prediction made for the initial film are independent of this extraction at the bottom.
The only consequence of the overestimated film thickness is the strong decrease of the tension, which reaches non-physical small values at the bottom meniscus.   

\begin{figure}[!ht]
  \centering
    \includegraphics[width=\linewidth]{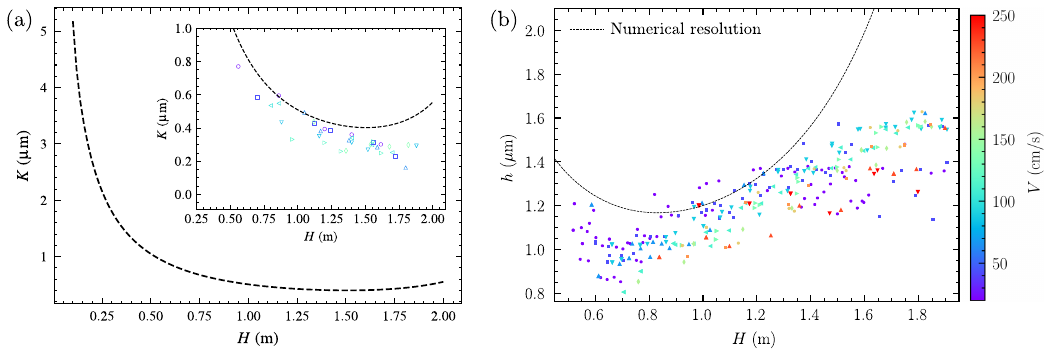}
  \caption{(a) \ER{Evolution of} the prefactor of the exponential profiles $K$ determined using the numerical resolution (black dashed line). \ER{The inset shows the comparison with the experimental data for different velocities (ranging from 20~cm/s to 140~cm/s). } (b) Evolution of the thickness $h$ of the soap films during generation as a function of the height $H$ reached by the films for all velocities probed between 20~cm/s and 250~cm/s. The colored points correspond to measurements taken at 30~cm above the liquid bath. These data are identical to those shown in the Fig. \ref{fig:spectro_master_curve} (b). The black dotted line corresponds to the evolution of $h$ given by the numerical resolution. }
   \label{Fig:comparaison_toutes-vitesses}
\end{figure}

In the following, we restrict the analysis to the initial film. 
The experimental and numerical  film profiles are  fitted by the exponential shape $K e^{(z-H)/\ell}$, and the obtained values of the prefactor $K$ are compared. They are plotted as a function of the film height in Fig. \ref{Fig:comparaison_toutes-vitesses} (a). 
Over the range of heights we can access in our experiment, between 0.5 and 1.9~m, the values of $K$ predicted by the numerical resolution are in agreement with our data \ER{(inset of Figure \ref{Fig:comparaison_toutes-vitesses})}. 
However, a discrepancy can be noticed when the height of the films becomes close to 2~m. This may be related to our method of redistributing the films in rectangular form, which becomes too far from the redistribution actually observed experimentally for the highest films.  

This observation is confirmed if we compare the measurements directly made by spectrometry, for example at 30~cm above the bath, with the numerical resolution. 
Indeed, in Fig. \ref{Fig:comparaison_toutes-vitesses} (b), we observe a good qualitative agreement between the model and the experiment. 
First a decrease of the thickness is noted, then a minimum value is obtained at a similar height of film, and finally a thickening is observed. 
However, this thickening is much faster than the increase in thickness measured experimentally. 

\section{Conclusion} 

The aim of this article is to describe the thickness profile of giant foam films during their generation. Our main result is the successful description of the thickness profile in the central part of the film. This necessitated a phenomenological description of the extraction of thick film from the lateral and bottom menisci. Our current understanding of these foam films generation is summarized below \ER{and occurs in two steps. First, there is a non-quasistatic step, not described in this manuscript, during which the initial film is formed. Then, in a second quasistatic step, this film entrained by the motion of the fishing line is both stretched upward and compressed on the sides by Frankel films extracted from the lateral menisci. It is also completed at the bottom by a Frankel film that continues to be extracted, this time in the quasistatic regime. Each of these physical processes still contains open questions.}


At short time, the liquid is entrained by viscosity around the wire, as it would be the case with a much more viscous liquid, like silicon oil \citep{champougny2017break}.
A reasonable hypothesis is that the reference state is prescribed by the moment at which this viscous flow is counterbalanced by the apparition of a strong enough Marangoni stress allowing entering the regime described in this article. \ER{This initial step determines the equilibrium shape of the film with a reference thickness $h_0$ and a reference area $A_0$. What fixes these values remains unknown.} \ER{An important open question is why the reference thickness is uniform and does not depend on the pulling velocity.  }

This initial film is then stretched and deformed, but keeps its exponential profile, predicted by the hydrostatic model.
This is true though the pulling velocities are large, up to meters per seconds. 
Additionally, the characteristic length of the exponential depends neither on the pulling velocity nor on the time. 
\ER{This confirms that the elastic model is a good description and that the memory of the initial step is conserved during the entire dynamics.}

\ER{We show that the extension of the initial film is mitigated by the extraction of thicker film from the lateral and the bottom menisci.}
The extraction law we use in the model is nevertheless not quantitative. 
As explained in the text, this can be due to the fact that we do not know the value of the surface tension in the menisci. 

Finally, when the motors stop, we expect a competition between the relaxation of the central film due to further extraction of thick films from the lateral meniscus and a thinning due to both drainage and evaporation. 
A quantitative description of these mechanisms would be necessary to predict the lifetime of the soap film. 

\ER{Our model catches the main physical mechanisms behind the generation of giant soap films.
The main limit of the model is that we do not describe quantitatively the extraction of thick film from the lateral and bottom meniscii. 
A better description necessitates a better understanding of the surfactants dynamics, which affects the surface tension in the mensicus.
Additionally, the exact geometry of the lateral film is not fully described. 
This would necessitate a better understanding of the liquid flow inside the thick Frankel's films. 
}

\ER{We would like to raise that the results obtained in this paper can actually be applied to much smaller soap films. 
The thin zone observed at the top of smaller films \cite{Saulnier2011} is probably also due to the extension of an initial film and our analysis, which should apply, predicts an exponential profile of the film thickness in this situation. 
What fixes the size of the initial film remains an open question.}

\section*{Acknowledgments}
We are grateful to Fran\c{c}ois Boulogne for the design of the Oospectro library, and Vincent Klein and Sandrine Mariot for the design of the experimental setup. 
Funding from ESA (MAP Soft Matter Dynamics)
and CNES (through the GDR MFA) is acknowledged. IC has received funding from the European Research Council (ERC) under the European Union’s Horizon 2020 research and innovation program (grant agreement No 725094).

\bibliographystyle{unsrtnat}
\bibliography{references}  






\end{document}